\documentclass[reprint,superscriptaddress,nofootinbib,amsmath,amssymb,aps,pra]{revtex4-1}
\usepackage{graphicx}
\usepackage{dcolumn}
\usepackage[inline]{enumitem}
\usepackage{bm}
\usepackage{amsmath}
\usepackage[font=small]{caption}
\usepackage[labelformat=empty, position=top]{subcaption}
\usepackage[export]{adjustbox}
\usepackage{blindtext}
\usepackage{amssymb}
\usepackage{newunicodechar}              

\begin{document}


\title{Robust Rhythmogenesis in the Gamma Band via Spike Timing Dependent Plasticity}

\author{Gabi Socolovsky}
  \email{gabisoco@post.bgu.ac.il}
  \affiliation{%
  	Department of Physics, Faculty of Natural Sciences, 
  }%
\affiliation{%
Zlotowski Center for Neuroscience, 
}
\author{Maoz Shamir}%

\affiliation{%
  	Department of Physics, Faculty of Natural Sciences, 
  }%
\affiliation{%
Zlotowski Center for Neuroscience, 
}

\affiliation{%
Department of Physiology and Cell Biology, Faculty of Health Sciences, Ben-Gurion University of the Negev, Be’er-Sheva, Israel
}


\date{\today}

\begin{abstract}
Rhythmic activity in the gamma band (30-100Hz) has been observed in numerous animal species ranging from insects to humans, and in relation to a wide range of cognitive tasks. Various experimental and theoretical studies have investigated this rhythmic activity. The theoretical efforts have mainly been focused on the neuronal dynamics, under the assumption that network connectivity satisfies certain fine-tuning conditions required to generate gamma oscillations. However, it remains unclear how this fine tuning is achieved.

Here we investigated the hypothesis that spike timing dependent plasticity (STDP) can provide the underlying mechanism for tuning synaptic connectivity to generate rhythmic activity in the gamma band.
We addressed this question in a modeling study. We examined STDP dynamics in the framework of a network of excitatory and inhibitory neuronal populations that has been suggested to underlie the generation of gamma. Mean field Fokker Planck equations for the synaptic weights dynamics are derived in the limit of slow learning. We drew on this approximation to determine which types of STDP rules drive the system to exhibit gamma oscillations, and demonstrate how the parameters that characterize the plasticity rule govern the rhythmic activity. Finally, we propose a novel mechanism that can ensure the robustness of self-developing processes, in general and for rhythmogenesis in particular. 

\end{abstract}

\maketitle


\section{\label{sec:Introducion}Introduction}
Rhythmic activity in the brain has been observed for more than a century \cite{jung1979fiftieth,buzsaki2006rhythms}. Oscillations in different frequency bands have been associated with different cognitive tasks and mental states \cite{buzsaki2006rhythms,buzsaki2015editorial,bocchio2017synaptic,shamir2009representation,taub2018oscillations}. Specifically, rhythmic activity in the Gamma band has been described in association with sensory stimulation \cite{gray1989oscillatory}, attentional selection \cite{fries2001modulation,bichot2005parallel}, working memory \cite{pesaran2002temporal} and other measures \cite{canolty2006high}. Deviation from normal rhythmic activity has been associated with pathology \cite{mably2018gamma,cao2018gamma,ghosh2013functional,uhlhaas2006neural}.

\begin{figure}[ht!]
\centering
\begin{subfigure}{0.2\textwidth}\captionsetup{justification=centering}

  \centering
  \includegraphics[width=3.5cm]{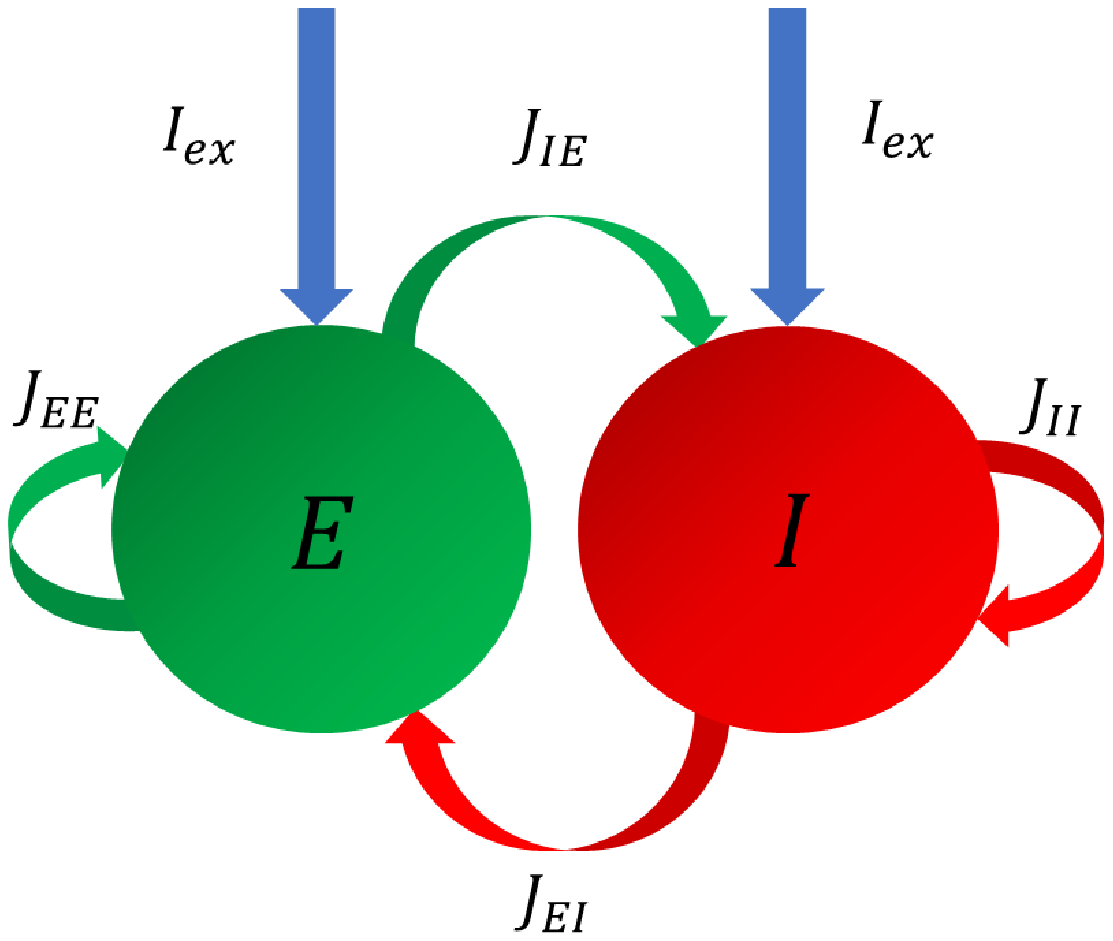}
  \begin{center}
      \caption{\textbf{(a)} \label{FigTwoPopulations}}
  \end{center}

\end{subfigure}\begin{subfigure}
{0.2\textwidth}\captionsetup{justification=centering}

  \centering
  \includegraphics[width=3.5cm]{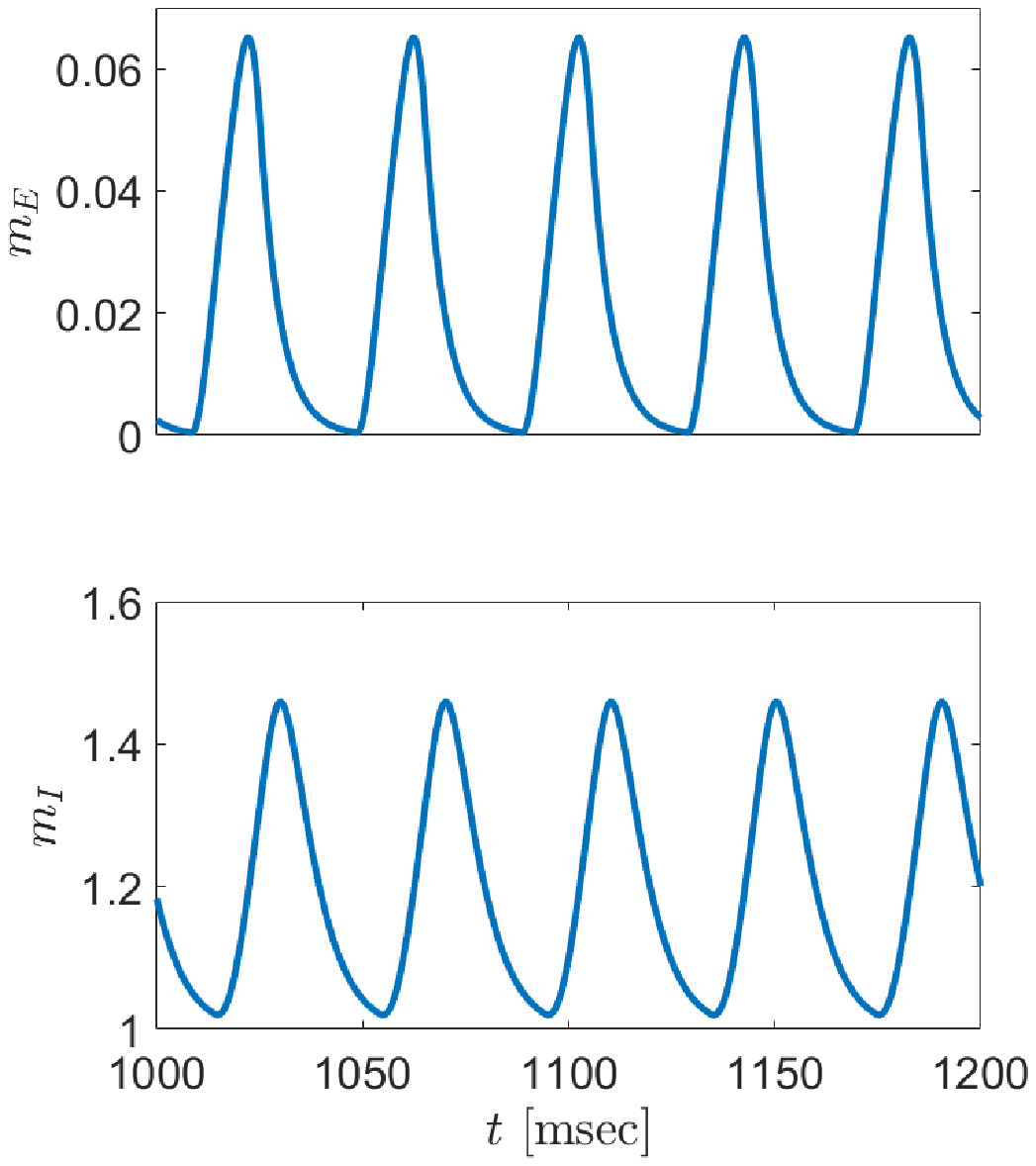}
  \begin{center}
      \caption{\textbf{(b)} \label{FigMeMiVsTime}}
  \end{center}
\end{subfigure}%
\hfill

\begin{subfigure}{0.2\textwidth}\captionsetup{justification=centering}

  \centering
  \includegraphics[width=3.5cm]{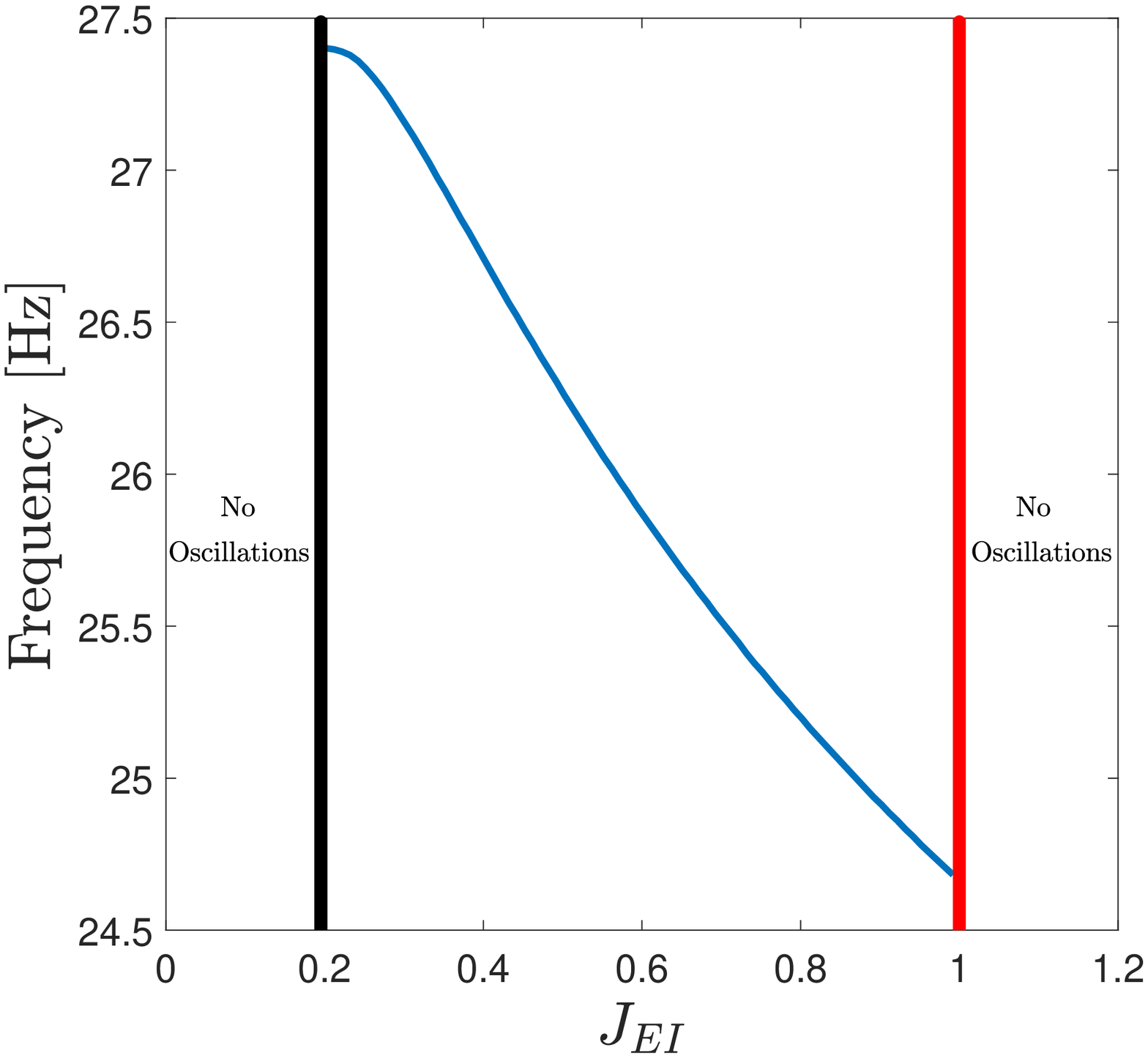}
  
  \begin{center}
      \caption{\textbf{(c)}\label{FigAreaOfOscillations}}
  \end{center}

\end{subfigure}\begin{subfigure}{0.2\textwidth}\captionsetup{justification=centering}

  \centering
  \includegraphics[width=3.5cm]{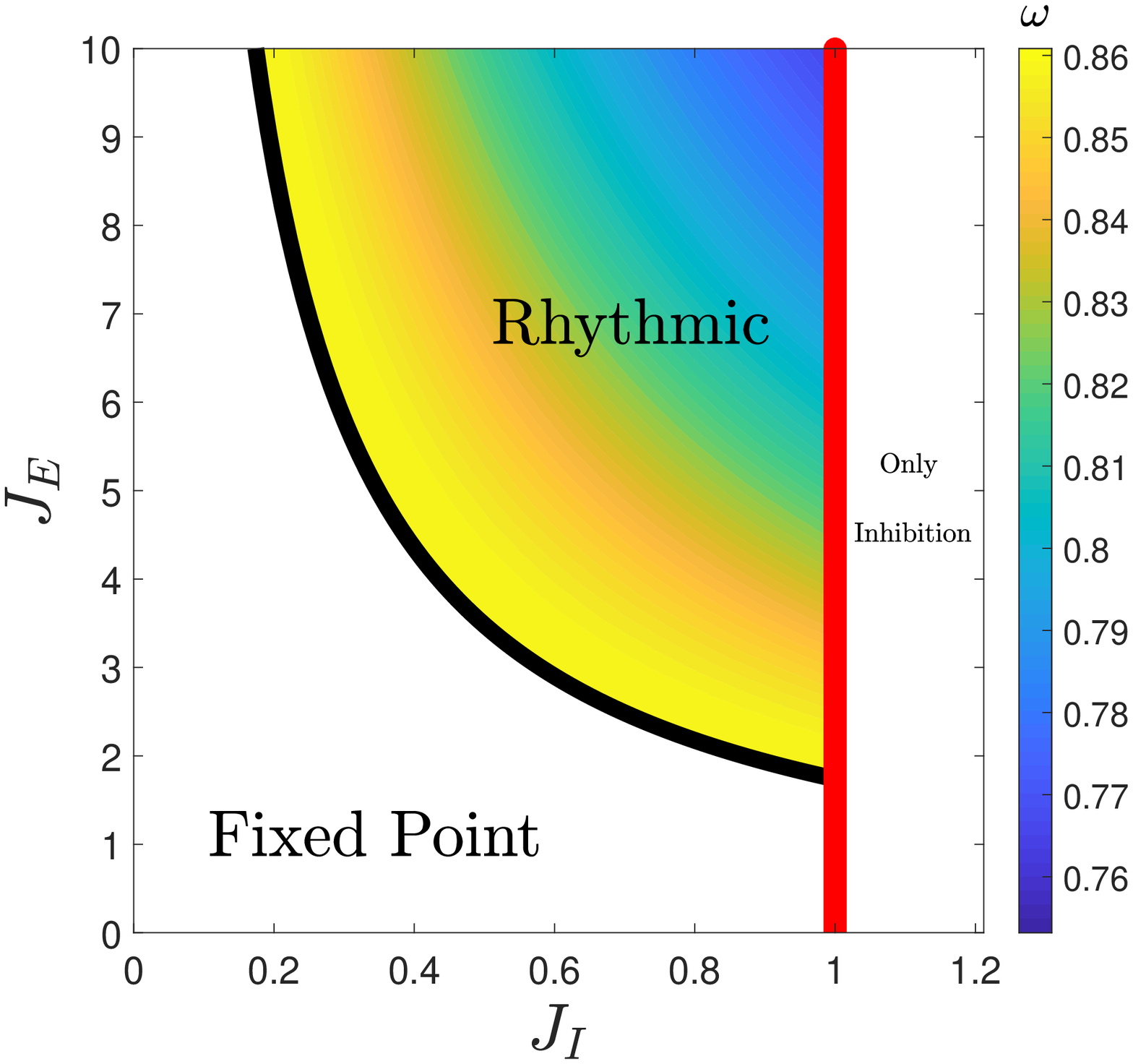}
  
  \begin{center}
      \caption{\textbf{(d)}\label{FigPhaseDiagram}}
  \end{center}

\end{subfigure}

\caption{The excitatory-inhibitory network.  ({\bf a}) Model architecture. The neuronal network here is composed of an excitatory (E) and an inhibitory (I) populations with inter- ($J_{IE}$ and $J_{EI}$) and intra- ($J_{II}$ and $J_{EE}$, that will be taken to zero hereafter) connections. The interaction is not symmetric and is delayed. ({\bf b}) The oscillatory dynamics of the mean excitatory end inhibitory population firing rates $m_E$ and $m_I$, respectively,  at a frequency of $24.9\text{Hz}$. Here we used  $J_{EE}=J_{II}=0$, $J_{IE}=8.91$, $J_{EI}=0.9$, $\tau_m = d = 5\text{ms}$. ({\bf c}) Rhythmic activity. The oscillation frequency is depicted as a function of the strength of the inhibitory to excitatory connection, $J_{EI}$. The following parameters were implemented here: $J_{EE}=J_{II}=0$, $\tau_m = d = 5\text{ms}$, $J_{IE}=8.91$.
({\bf d}) Phase diagram of the delayed rate model. Strong inhibition, $J_I>1$, leads the system to a purely inhibitory fixed point, in which $m_E$ is fully suppressed, and $m_I=1$. For weak to moderate inhibition, $J_{I}\in\left(0,1\right)$, and below the black line, the system converges to a fixed point, where both populations are active. For $\bar{J}>\bar{J}_d$ (above the black line) the model exhibits rhythmic activity. For a given time delay, the frequency is governed by $\bar{J}$. When $\bar{J}$ is increased, higher frequencies are observed. Here we used $\tau_m=d=1$.}

\label{FigArchiterctureAndFreuqencyVsJ}
\end{figure}

Considerable theoretical efforts have been devoted to unraveling the neural mechanism responsible for generating rhythmic activity in the Gamma band \cite{fries2007gamma,amir2018vigilance,welle2017new,womelsdorf2006gamma,luz2016oscillations,roxin2006rate}. One possible mechanism is based on delayed inhibitory feedback \cite{roxin2006rate,battaglia2007temporal,soloduchin2018rhythmogenesis,battaglia2011synchronous}. The basic architecture of this mechanism is composed of one excitatory and one inhibitory neuronal populations, with reciprocal connections (Fig. \ref{FigTwoPopulations}). A target rhythm is obtained by tuning the strengths of the excitatory and inhibitory interactions (Fig. \ref{FigMeMiVsTime}-\ref{FigAreaOfOscillations}). However, it is unclear which mechanism results in the required fine-tuning \cite{soloduchin2018rhythmogenesis,shamir2019theories}.

We hypothesized that activity dependent synaptic plasticity can provide the mechanism for tuning the interaction strengths in order to stabilize a specific rhythmic activity in the gamma band. 

Here we focused on spike timing dependent plasticity (STDP) as the rhythmogenic process \cite{soloduchin2018rhythmogenesis,shamir2019theories}. Below, we briefly describe STDP and derive the dynamics of the synaptic weights in the limit of slow learning. Since the cross-correlation of neural activity is central to STDP dynamics, we next define the network dynamics and analyze its phase diagram and the dependence of the correlations on the synaptic weights. Using the separation of timescales in the limit of slow learning we analyze STDP dynamics and investigate under what conditions STDP can stabilize a specific rhythmic activity and how the characteristics of the STDP rule govern the resultant rhythmic activity. Finally, we summarize our results, discuss possible extensions and limitations and propose a general principle for robust rhythmogenesis.

\section{\label{sec:STDP}STDP dynamics}
The basic coin of information transfer in the central nervous system is the spike: a short electrical pulse that propagates along the axon (output branch) of the transmitting neuron to synaptic terminals that relay the information to the dendrites (input branch) of the receiving neurons downstream. While spikes are stereotypical, the relayed signal depends on the synaptic weight, which can be thought of as interaction strength. Learning is the process that modifies synaptic weights (the dynamics of the interaction strengths themselves), and typically occurs on a slower timescale than the timescale of the neuronal responses. 

STDP is an empirically observed microscopic learning rule in which the modification of the synaptic weight depends on the temporal relation between the spike times of the pre (transmitting) and post (receiving) synaptic neurons \cite{cateau2003stochastic,gutig2003learning,morrison2008phenomenological,rubin2001equilibrium,song2001cortical}. Following \cite{luz2014effect} the change, $\Delta J_{ ij }$, in the synaptic weight $J_{ ij }$ from the pre-synaptic neuron $j$ to the post-synaptic neuron $i$ is expressed as the sum of two processes: potentiation (i.e., increasing the synaptic weight) and depression (i.e., decreasing the synaptic weight):
\begin{equation}
\label{eq:DelatJ}
    \Delta J_{ij} = \lambda \left[ K_+(\Delta t) - \alpha K_-(\Delta t) \right],
\end{equation}
where $ \Delta t = t_{i} - t_{j}$ is the pre and post spike time difference. Functions $K_{ \pm} (t) \geq 0$ describe the temporal structure of the potentiation ($+$) and depression ($-$) of the STDP rule. Parameter $\alpha$ denotes the relative strength of the depression and $\lambda$ is the learning rate. We assume that learning occurs on a slower timescale than the characteristic timescales that describe neuronal activity (for given fixed synaptic weights). 

A wide range of temporal structures of STDP rules has been reported \cite{bi1998synaptic,woodin2003coincident,suvrathan2016timing,haas2006spike,bell1997synaptic,nishiyama2000calcium,froemke2005spike,tzounopoulos2007coactivation}. Here we focus on two families of rules. One is composed of temporally symmetric rules and the other is made up of temporally asymmetric rules.

For the temporally symmetric family we apply a difference of Gaussian STDP rule; namely:
\begin{equation}
    \label{eq:DOG}
    K_{ \pm} (t) = \frac{1}{ \sqrt{2 \pi } \tau_{ \pm} } e^{ - ( t/\tau_{ \pm} )^2 /2} ,
\end{equation} 
where $\tau_{ \pm}$ denotes the characteristic time scales of the potentiation ($+$) and depression ($-$). Consistent with the popular description of the famous Hebb rule that `neurons that fire together wire together' \cite{hebb1961organization}, we refer to the case of $\tau_+ < \tau_-$ as Hebbian, and $\tau_+ < \tau_-$ anti-Hebbian.

For the temporally a-symmetric STDP rules we take:
\begin{equation}
    \label{eq:TA}
    K_{ \pm} (t) = \frac{1}{\tau_{ \pm}} e^{\mp H t/\tau_{ \pm}} \Theta(\pm H t)
    ,
\end{equation} 
where $\Theta (x)$ is the Heaviside step function, $\tau_{ \pm}$ are the characteristic time scales of the potentiation (+) and depression ($-$), and $H = \pm 1$ dictates the Hebbianity of the STDP rule. The rule will be termed Hebbian for $H = 1$, when potentiation occurs in the causal branch, $t_{\mathrm{post}} > t_{\mathrm{pre}}$, and anti-Hebbian for $H=-1$. 

Different types of synapses have been reported to exhibit different types of STDP rules. Consequently, there is no a-priori reason to assume that excitatory and inhibitory synapses share the exact same learning rule. In particular, the characteristic time constants $\tau_{E, \pm}$ for excitatory synapses and $\tau_{I, \pm}$ for inhibitory synapses may differ. 


Changes to synaptic weights due to the plasticity rule of Eq.~(\ref{eq:DelatJ}) at short time intervals occur as a result of either a pre or post-synaptic spike during this interval. Thus,
\begin{eqnarray}
\label{eq:dJ_dt1}
    \dot{J}_{i,j} (t) & = & \lambda \rho_{i} (t) \int^{ \infty}_0 \rho_{j}(t-t')
    \left[K_+(t') - \alpha K_-(t') \right] dt' \\
    \nonumber
    &+&
    \lambda \rho_{j} (t) \int^{ \infty}_0 \rho_{i}(t-t')
    \left[ K_+(-t') - \alpha K_-(-t') \right] dt'
    ,
\end{eqnarray}
where $\rho_{ \mathrm{post / pre} } (t) = \sum_l \delta (t - t_l^{\mathrm{post / pre}} )$ is the spike train of the post/pre neuron written as the sum of the delta function at the neuron's spike times $\{ t_l^{\mathrm{post / pre}} \}_l $. In the limit of slow learning, $\lambda \rightarrow 0$, the right hand side of Eq.~(\ref{eq:dJ_dt1}) can be replaced by its temporal mean (see \cite{luz2014effect} for complete derivations). This approximation, has been termed the mean field Fokker-Planck approximation \cite{gutig2003learning}. As fluctuations vanish in this limit and deterministic dynamics are retained for the mean synaptic weights, we get
\begin{eqnarray}
    \label{eq:dJ_dt2}
    \dot{J}_{ij} (t) & = & \lambda \int_{ - \infty}^{ \infty} \Gamma_{ij}(- t')
    \left[K_+(t') - \alpha K_-(t') \right] dt'
    ,
\end{eqnarray}
where $\Gamma_{ij}( t )$ is the cross correlation of neurons $i$ and $j$:
\begin{equation}
    \Gamma_{ij}( t) = \langle \rho_i(t') \rho_j(t' + t) \rangle
    .
\end{equation}
The angular brackets $\langle \cdots \rangle $ denote ensemble averaging over the neurnal noise and temporal averaging over one period in the case of rhythmic activity (see Soloduchin \cite{soloduchin2018rhythmogenesis} \& Shamir  and Luz \& Shamir \cite{luz2014effect} for more details).
Note that the dependence of the r.h.s.\ of Eq.~(\ref{eq:dJ_dt2}) on time, $t$, occurs through the dependence of the cross-correlations on the synaptic weights at time $t$.
Thus, the key to analyzing STDP dynamics is the ability to compute the cross-correlations of the neural activities and grasp their dependence on the synaptic weights. To this end we examined rhythmogenesis in the gamma band using the framework of a reduced rate model with delay, proposed by Roxin and colleagues ({Fig. \ref{FigTwoPopulations}}). A complete analysis of the model appears in \cite{roxin2006rate,battaglia2011synchronous,battaglia2007temporal}. Below we briefly describe the phase diagram of the system and derive the cross-correlations.

\section{\label{sec:PhaseDiag}The delayed excitatory inhibitory network}
The firing of different neurons is assumed to follow independent inhomogeneous Poisson process statistics with instantaneous firing rates that adhere to the reduced model in Roxin et al. \cite{roxin2006rate}. In their work, Roxin and colleagues considered a full model that included inter-population as well as intra-population interactions (i.e., excitatory-excitatory and inhibitory-inhibitory). Here, for simplicity we restrict the analysis to the minimal model that can reproduce oscillations in the gamma band. To do so, we model the rate dynamics of the gamma generating network by:
\begin{eqnarray}\label{eqn:firing rates dynamics1}
    \tau_m \dot{m}_E (t) &=& - m_E(t) + [I - J_{I} m_I(t-d) ]_+
    \\\label{eqn:firing rates dynamics2}
    \tau_m \dot{m}_I (t) &=& - m_I(t) + [I + J_{E} m_E(t-d) ]_+,
\end{eqnarray}
where $m_{E/I} (t)$ is the mean firing rate of the excitatory (E) and inhibitory (I) population at time $t$. $\tau_m$ is the neuronal time constant. Unless stated otherwise, we take $\tau_m = 1$, which is equivalent to measuring time in units of the neuronal time constant. Parameter $d$ denotes the delay, and $I$ is the external input to the system. In our analysis we took $I=1$. $J_{E}$ and $J_{I}$ are the effective interaction strengths between the two populations. $J_{E}$ ($J_{I}$) can be thought of as a global order parameter reflecting the mean synaptic weight from the excitatory (inhibitory) pre-synaptic population to the inhibitory (excitatory) post-synaptic population.

For strong inhibition, $J_I>1$, the system converges to a fixed point in which the excitatory population is fully suppressed by the inhibitory population, $\vec{m}^*=\begin{pmatrix} 0\\ 1
\end{pmatrix}$. 
For weak to moderate levels of inhibition, $J_I \in (0,1)$, the system has a fixed point in which both populations are active, $\vec{m}^* \equiv \begin{pmatrix}m_E^* \\m_I^* \end{pmatrix} = \frac{1}{1+\bar{J}^2} \begin{pmatrix} 1-J_I \\ 1+J_E
\end{pmatrix}$, with $\bar{J}\equiv \sqrt{J_E J_I}$. 
However, this fixed point is not stable for  $\bar{J} > \bar{J}_d$, where $\bar{J}_d^2=1+\omega_d^2$,  $\omega_d=\cot\left(\omega_d d\right)$ and $\omega_d\in \left[0,\pi/2d\right]$ (see Roxin et al. \cite{roxin2006rate}). In this region ($\bar{J} > \bar{J}_d$ and $J_I <1$) the system converges to a limit cycle solution, Fig.\ \ref{FigPhaseDiagram}.

\begin{figure*}[ht]
    \centering
    \begin{subfigure}[b]{0.3\textwidth}\captionsetup{justification=centering}
    \centering
        \includegraphics[width=5cm]{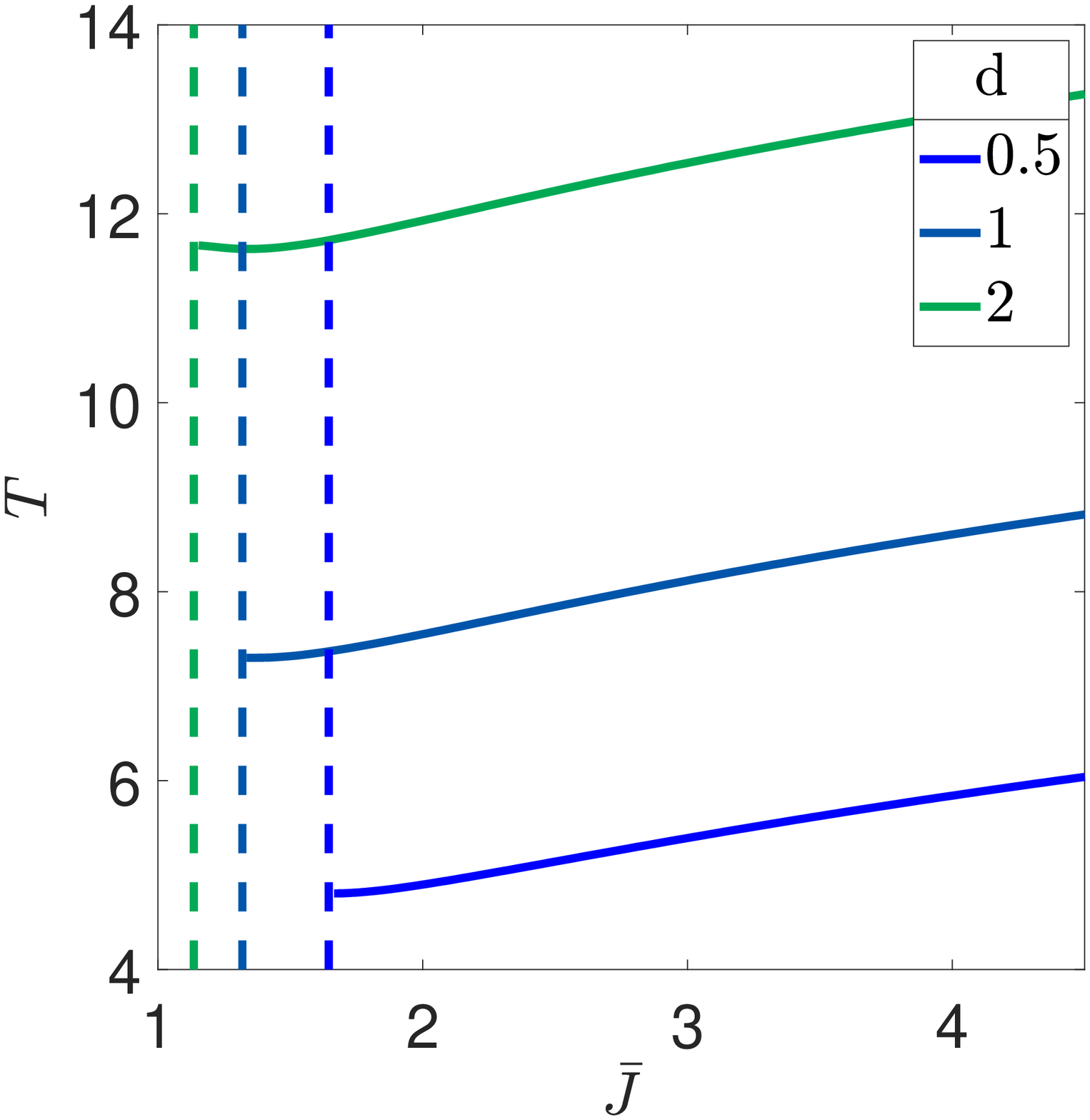}
        \begin{center}
            \caption{\textbf{(a)}\label{FigXStatsPeriod}}
        \end{center}
        
    \end{subfigure}\quad \begin{subfigure}[b]{0.3\textwidth}\captionsetup{justification=centering}
        \centering
        \includegraphics[width=5cm]{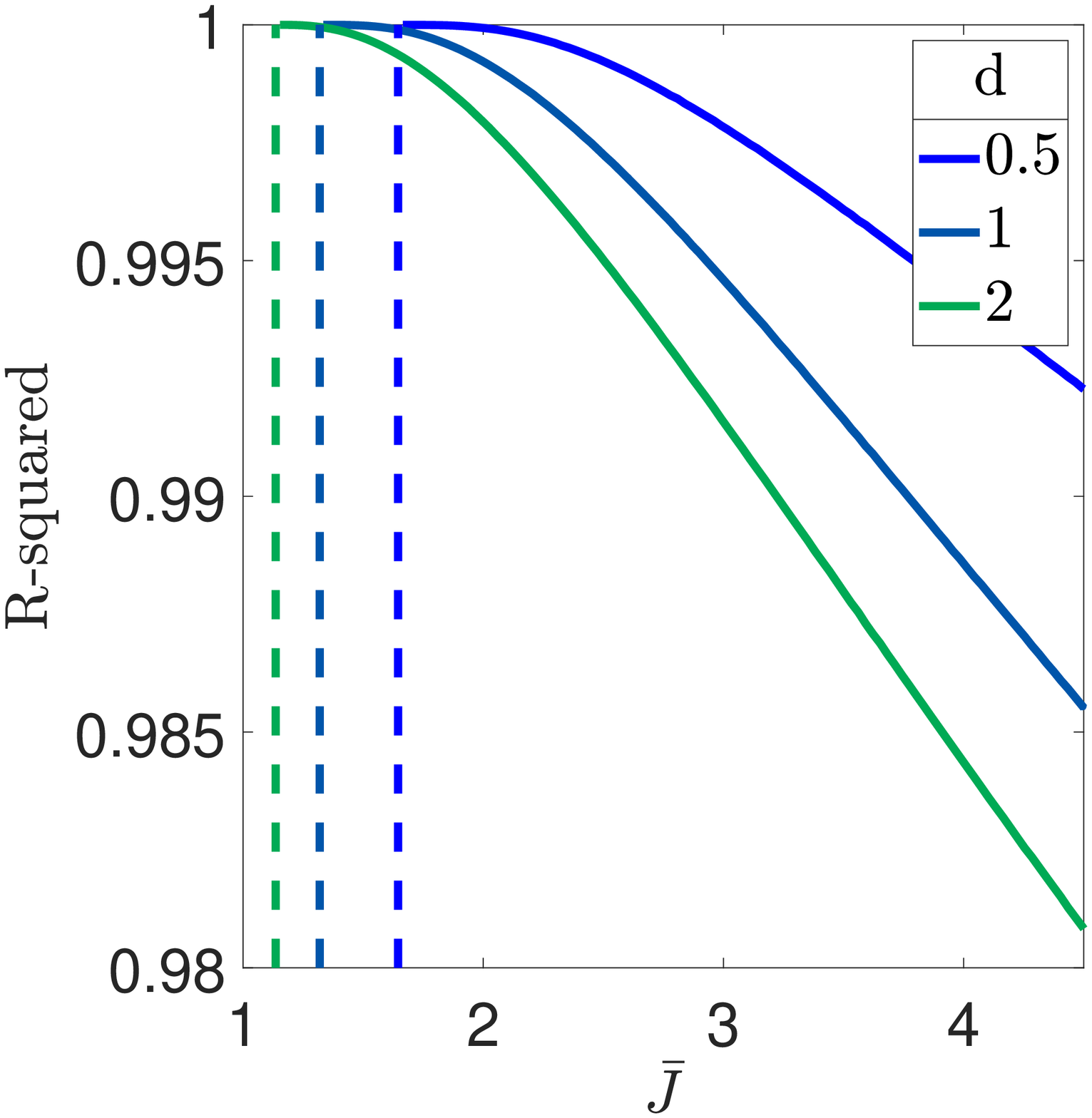}
         \begin{center}
            \caption{\textbf{(b)}\label{FigXStatsRSquared}}
        \end{center}
              
    \end{subfigure}\quad  \begin{subfigure}[b]{0.3\textwidth}\captionsetup{justification=centering}
        \centering
        \includegraphics[width=5cm]{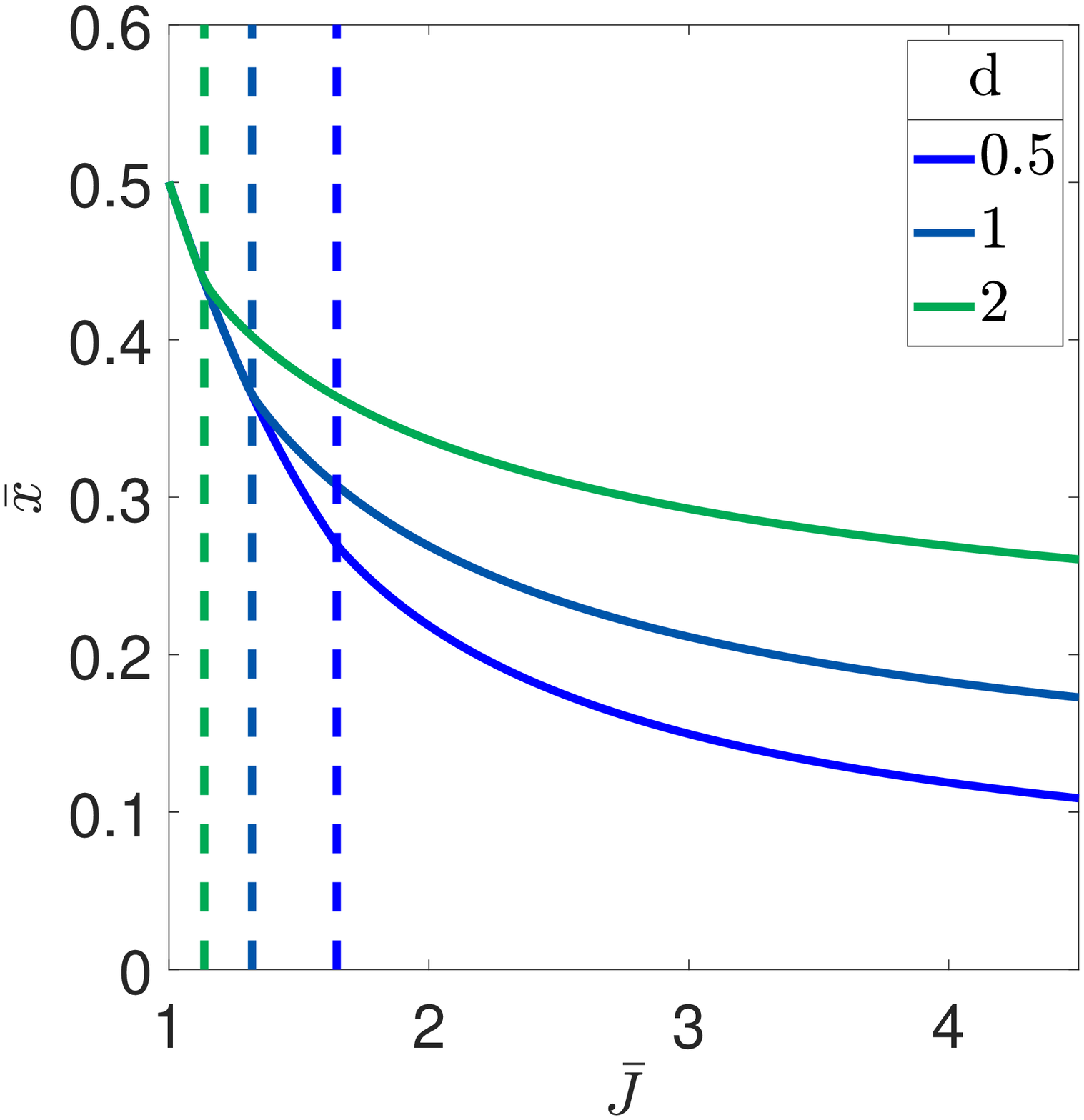}
        \begin{center}
            \caption{\textbf{(c)}\label{FigXStatsXBar}}
        \end{center}
        
    \end{subfigure}
    
    \begin{subfigure}[b]{0.3\textwidth}\captionsetup{justification=centering}
        \centering
        \includegraphics[width=5cm]{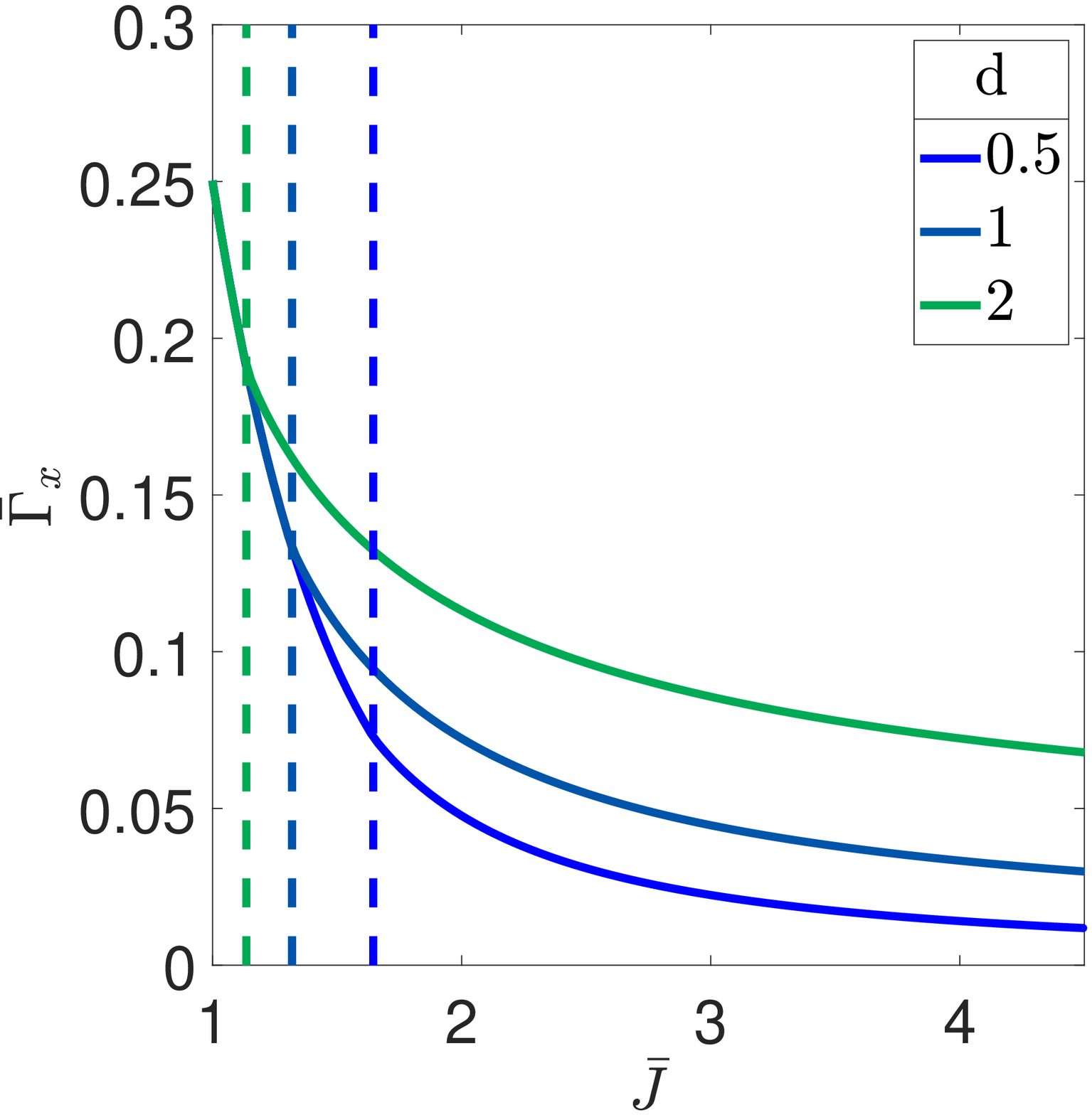}
        \begin{center}
            \caption{\textbf{(d)}\label{FigXStatsGammaXbar}}
        \end{center}
        
    \end{subfigure}\quad \begin{subfigure}[b]{0.3\textwidth}\captionsetup{justification=centering}
        \centering
        \includegraphics[width=5cm]{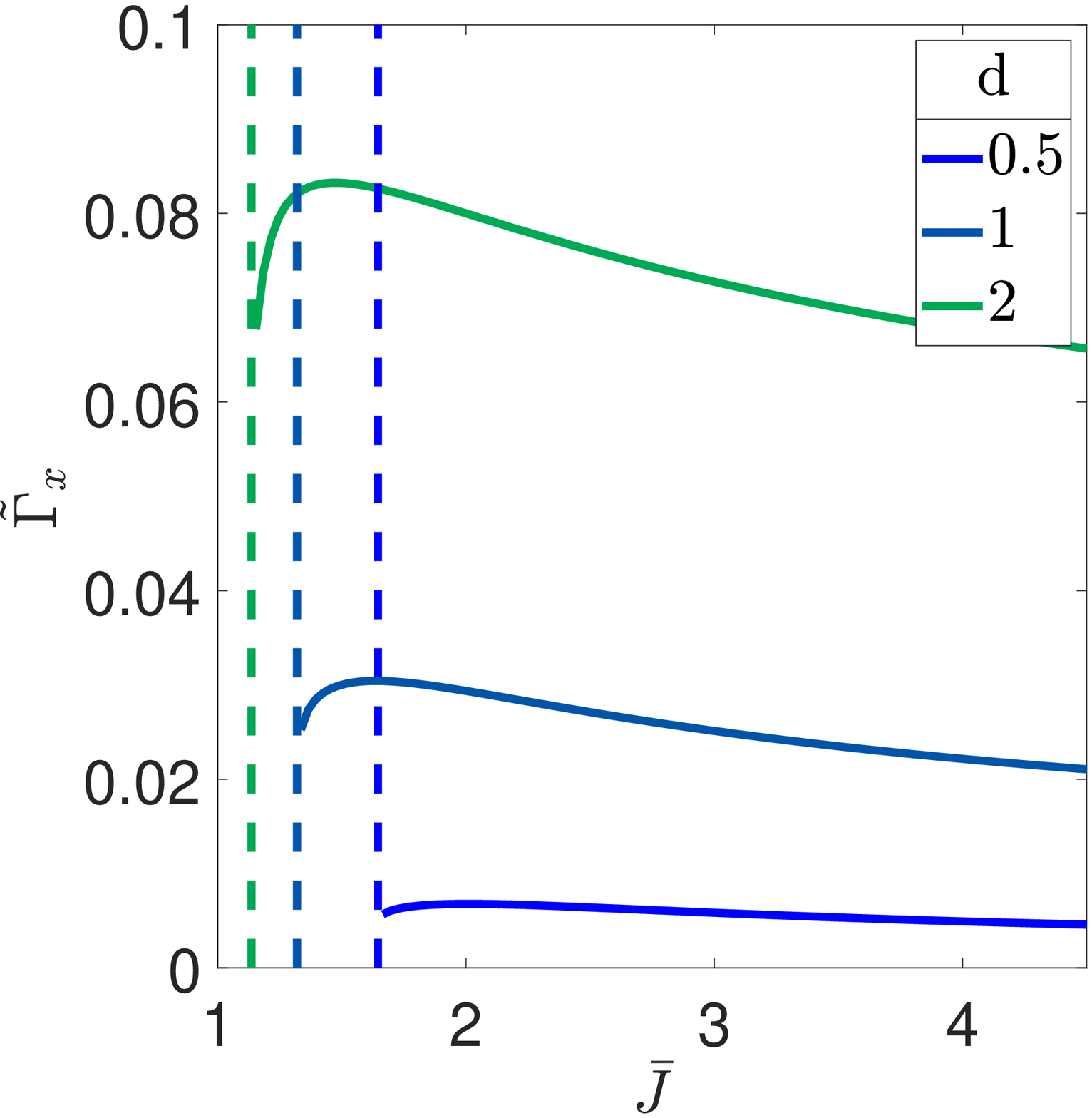}
         \begin{center}
            \caption{\textbf{(e)}\label{FigXStatsGammaXTilde}}
        \end{center}
        
        \end{subfigure}
    \caption{Dynamics of the one dimensional variable $x$, Eq.~(\ref{eq:x_dot}). Different features characterizng the dynamics of $x$ are shown as a function of $\bar{J}$, for different values of the delay, $d$, differentiated by color. The dashed lines indicate the values of $\bar{J}_d$, for different delays, shown by color. The parameters plotted are:   ({\bf \protect\subref{FigXStatsPeriod}}) The period of the oscillations. ({\bf \protect\subref{FigXStatsRSquared}}) The goodness of fit of the cosine approximation, Eq.~(\ref{eq:CosApprox}), of $\Gamma_x$. $R^2$. ({\bf \subref{FigXStatsXBar}}) The temporal average of $x$,  $\bar{x}$. ({\bf  \subref{FigXStatsGammaXbar}}) The zeroth order Fourier component of the auto-correlation of $x$, $\bar{\Gamma}_x$. ({\bf \subref{FigXStatsGammaXTilde}}) The first order Fourier component of the auto-correlation of $x$, $\tilde{\Gamma}_x$.}\label{fig:x-stats}.

    ~ 
    
    ~ 
   
\end{figure*}

By rescaling the firing rates, the two dimensional first order delayed dynamics, Eq.~(\ref{eqn:firing rates dynamics1})-(\ref{eqn:firing rates dynamics2}), can be reduced to a one dimensional delayed dynamic equation: 
\begin{equation}
\label{eq:x_dot}
    \ddot{x}(t) + 2\dot{x}(t) + x(t) - [1-\bar{J}^2 x(t-2d) ]_+ = 0
    ,
\end{equation}
with
\begin{eqnarray}
\label{eq:mi_ansatz}
    m_I(t) &=& 1+\left(J_E-\bar{J}^2\right)x\left(t\right)
    \\
\label{eq:me_ansatz}
    m_E(t-d) &=& 
    \frac{J_E-\bar{J}^2}{J_E}\left(x\left(t\right)+\dot{x}\left(t\right)\right)
    .
\end{eqnarray}
Equation~(\ref{eq:x_dot}) highlights the fact that the temporal structure of the limit cycle solution depends on the synaptic weights, $J_E$ and $J_I$, only via $\bar{J}$. In particular, the period of oscillations is solely a function of $\bar{J}$ and $d$.
As shown in Fig.\ \ref{FigXStatsPeriod} the period is a monotonically increasing function of both $\bar{J}$ and $d$.

In our model the cross-correlations are given by temporal averaging of the mean firing rates, $\Gamma_{IE}\left(\Delta\right)\equiv \langle m_I\left(t\right) m_E\left( t+\Delta\right)\rangle$. In the fixed point region of the phase diagram, the cross-correlation is simply given by the product of the mean rates, $\Gamma_{IE}\left(\Delta\right)=\left(1-J_I\right)\left(1+J_E\right)/\left(1+\bar{J}^2\right)^2$. 

In the region of the phase diagram where the system converges to a limit cycle solution, Eqs.~(\ref{eq:mi_ansatz}) \& (\ref{eq:me_ansatz}) provide the scaling of the cross-correlations; namely,
\begin{equation}\label{eqn:Cross-correlation_first}
    \Gamma_{IE}\left(\Delta\right)=a\bar{x}+b\left(\Gamma_x\left(\Delta+d\right)+\frac{d}{d\Delta}\Gamma_x\left(\Delta+d\right)\right)
\end{equation}
where $a\equiv\left(J_E-\bar{J}^2\right)/J_E$, $b \equiv \left(J_E-\bar{J}^2\right)^2/J_E$, $\bar{x}\equiv \langle x\left(t\right)\rangle$, and $\Gamma_x\left(s\right)\equiv \langle x\left(t\right)x\left(t+s\right)\rangle$. Note that $\bar{x}$ and $\Gamma_x$ depend solely on  the delay, $d$, and $\bar{J}$. Numerical investigation reveals that the auto-correlation of $x(t)$ is well approximated by a cosine function,
\begin{eqnarray}
\label{eq:CosApprox}
    \Gamma_x\left(s\right)\approx \bar{\Gamma}_x +\tilde{\Gamma}_x \cos\left(\omega s\right)
\end{eqnarray}
where we used
\begin{eqnarray}
    \bar{\Gamma}_x &=& \int_0^T \Gamma_x\left(s\right) ds/T 
    \\
    \tilde{\Gamma}_x &=& 2 \int_0^T \Gamma_x\left(s\right) \cos (2 \pi s/T) ds/T
    ,
\end{eqnarray}
with $T$ denoting the period of the limit cycle, as can be seen from the value of $R^2$, Fig.\ \ref{FigXStatsRSquared}. The goodness of fit of the cosine approximation decreases when $\bar{J}$ or $d$ are increased. Nevertheless, for a wide range of parameters relevant to the generation of gamma oscillations $R^2$ is extremely high ($R^2 > 0.98$ throughout Fig.\ \ref{FigXStatsRSquared}). Both $\bar{x}$ and $\bar{\Gamma}_x$ monotonically decrease as $\bar{J}$ increases, Fig.\ \ref{FigXStatsXBar}-\ref{FigXStatsGammaXbar}. In addition, they transition the bifurcation line continuously. On the other hand, $\tilde{\Gamma}_x$ does not transition in a continuous manner: it is zero in the fixed point region and jumps to a positive value in the rhythmic region, Fig.\ \ref{FigXStatsGammaXTilde}. 

\begin{figure}[ht!]
    
    \includegraphics[width=8.6cm]{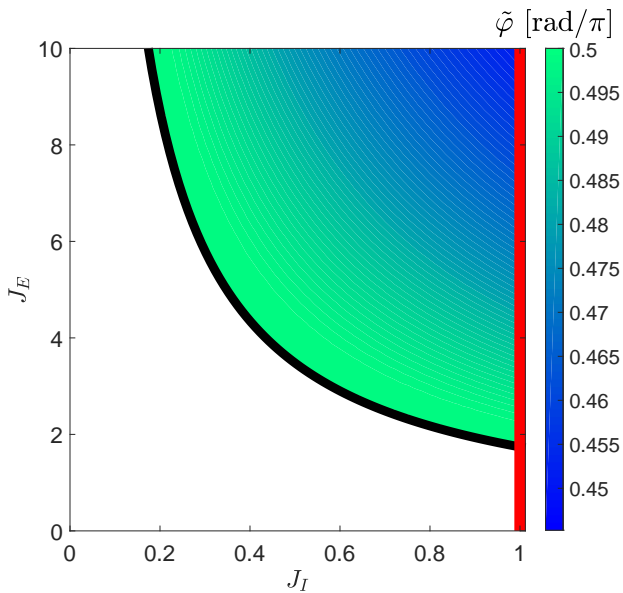}
    \caption{The values of $\tilde{\varphi}$. The parameter, $\tilde{\varphi} / \pi$ is shown by color as a function of $J_E$ and $J_I$ on the phase diagram. Here $d=1$ was used.}
    \label{fig:PhiTilde}
\end{figure}

Using the cosine approximation for the correlations, Eq.~(\ref{eq:CosApprox}) and the scaling Eqs.~(\ref{eq:mi_ansatz}) \& (\ref{eq:me_ansatz}) yields the semi-empirical excitatory-inhibitory cross-correlations function:
\begin{equation}
\label{eq:SemiEmpirical}
    \Gamma_{IE}\left(\Delta\right)\approx \bar{\Gamma}+\tilde{\Gamma}\cos\left(\omega\Delta+\tilde{\varphi}\right)
\end{equation}
with $\bar{\Gamma}=a\bar{x}+b\bar{\Gamma}_x$, $\tilde{\Gamma}=b\tilde{\Gamma}_x\left(1+\omega^2\right)^{1/2}$, $\tilde{\varphi}=\omega\left(d+\varphi_\omega\right)$ and $\omega\varphi_\omega=\arcsin\left(\omega/\sqrt{1+\omega^2}\right)$. Figure \ref{fig:PhiTilde} shows the values of $\tilde{\varphi}$  on the phase diagram. The phase, $\tilde{\varphi} ( \bar{J}, d)$ is $\pi/2$ on the bifurcation line and weakly decreases as $\bar{J}$ is further increased.  
Note that $\Gamma_{EI}\left(\Delta\right)=\Gamma_{IE}\left(-\Delta\right)$. 

\section{\label{sec:Flow}STDP induced flow on the phase diagram}
Utilizing the semi-empirical cross-correlations, Eq.~(\ref{eq:SemiEmpirical}), yields the following dynamics for the synaptic weights:
\begin{equation}
\label{eq:StdpDyn}
    \dot{J}_{\sigma}=\lambda\left(\bar{\Gamma}\bar{K}+\tilde{\Gamma}\tilde{K}_\sigma\right)    
\end{equation}
where
\begin{equation}
    \bar{K}=\int_{-\infty}^{\infty}{K\left(\Delta\right)d\Delta},
\end{equation}
\begin{equation}
    \tilde{K}_\sigma\equiv \int_{-\infty}^{\infty}{K\left(\Delta\right)\cos(\omega\Delta_\sigma+\tilde{\varphi})d\Delta},
\end{equation}
where $\sigma=E,I$, we used the notation $K\left(\Delta\right)=K_+\left(\Delta\right)-\alpha K_-\left(\Delta\right)$, and $\Delta_E=-\Delta$ whereas $\Delta_I=\Delta$.

The STDP dynamics, Eq.~(\ref{eq:StdpDyn}), induces a flow in the phase plane of the synaptic weights $[J_I, J_E]$, which is also the phase diagram of the neural responses. The right hand side of Eq.~(\ref{eq:StdpDyn}) depends on the synaptic weights through the Fourier transforms of the cross-correlations $\bar{\Gamma}$ and $\tilde{\Gamma}$. Using the separation of timescales between the fast neuronal responses and the slow learning rate, in the limit of slow learning $\lambda \rightarrow 0$, one can compute $\bar{\Gamma}$ and $\tilde{\Gamma}$ from the neuronal dynamics for fixed synaptic weights.

Thus, STDP induces a flow on the phase diagram of the system. Rhythmogenesis is obtained when this flow guides the system and stabilizes it at a fixed point on the phase diagram that is characterized by the desired rhythm.

In the region of the phase diagram in which the mean neuronal firings relax to a fixed point, $\tilde{\Gamma} =0$. Due to our choice of normalization, $\bar{K}_\pm=1$, in this region $ \mathrm{sign} ( \dot{J}_{\sigma} ) = \mathrm{sign} (1 - \alpha_\sigma) $, $\sigma \in \{ E, I\} $. Consequently,  the STDP dynamics will induce a flow from the fixed point region towards the rhythmic region if and only if the potentiation is strong relative to the depression for both types of synapses, $\alpha_E, \alpha_I < 1$ (except for a small region of the phase diagram with high inhibition and low excitation, which also depends on the learning rates, $\lambda_E$ and $\lambda_I$, of the different synapses). This result holds true for any STDP rule. 
In contrast, the STDP dynamics in the rhythmic region of the phase diagram depend on the temporal structure of the learning rule.

\begin{figure*}[ht!]
    \centering
    \begin{subfigure}[b]{0.3\textwidth}\captionsetup{justification=centering}
        \includegraphics[width=5cm]{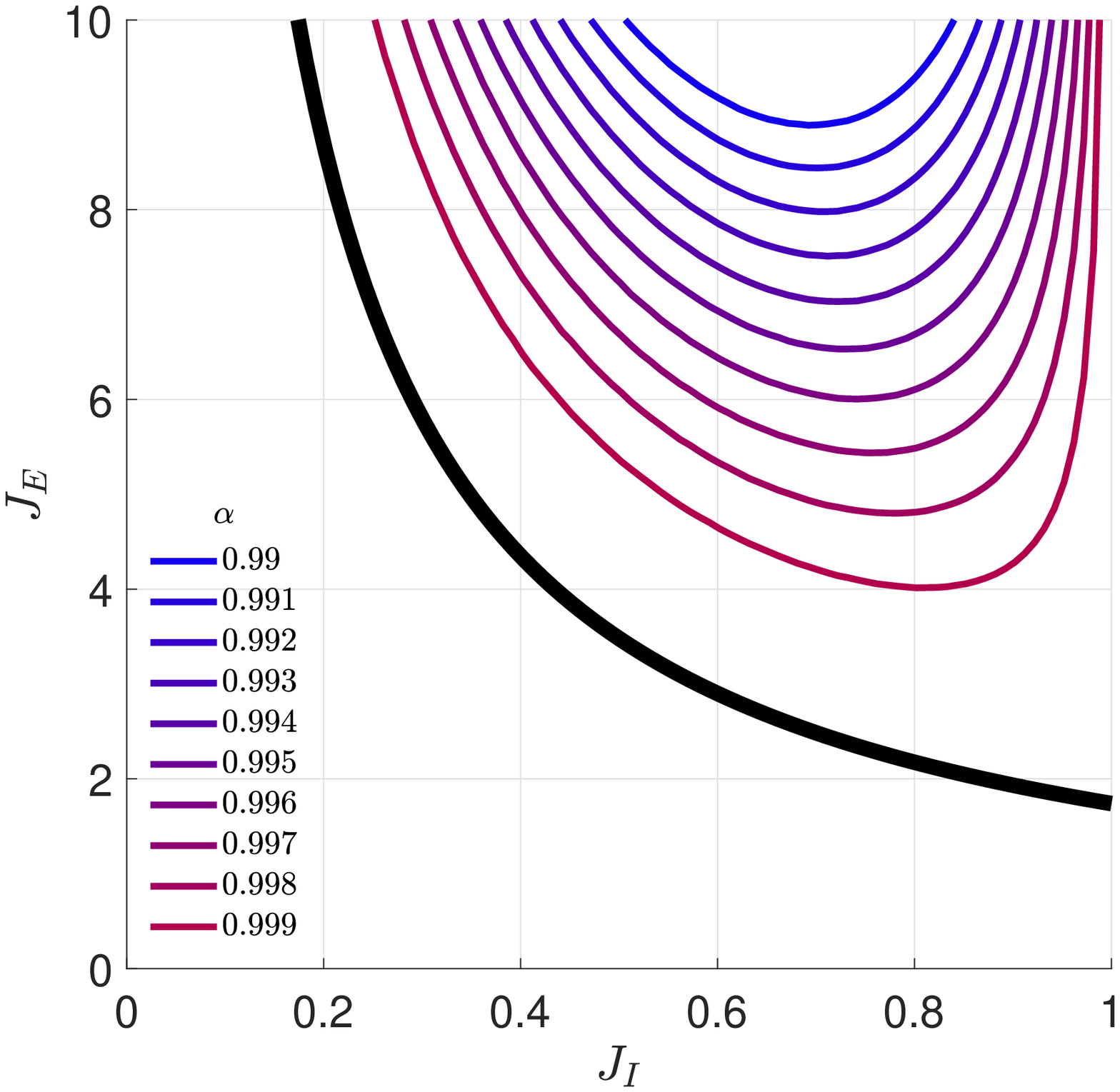}
        \begin{center}
            \caption{\textbf{(a)} \label{fig:gauss_alpha_change}}
        \end{center}
       
    \end{subfigure}\quad \begin{subfigure}[b]{0.3\textwidth}\captionsetup{justification=centering}
        \includegraphics[width=5cm]{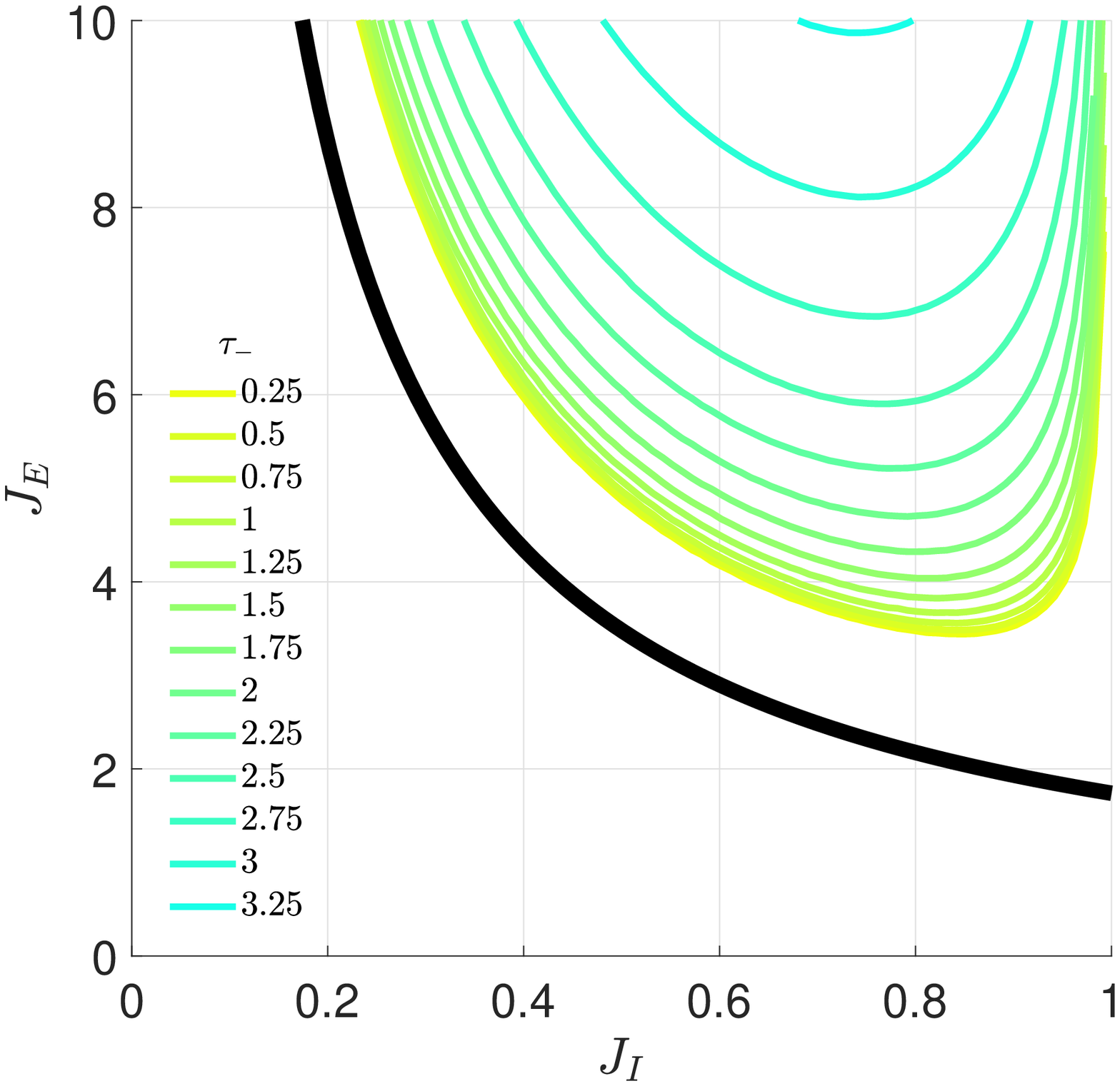}
       \begin{center}
            \caption{\textbf{(b)}  \label{fig:gauss_tau_change}}
        \end{center}
       
    \end{subfigure}
    
    \begin{subfigure}[b]{0.3\textwidth}\captionsetup{justification=centering}
        \includegraphics[width=5cm]{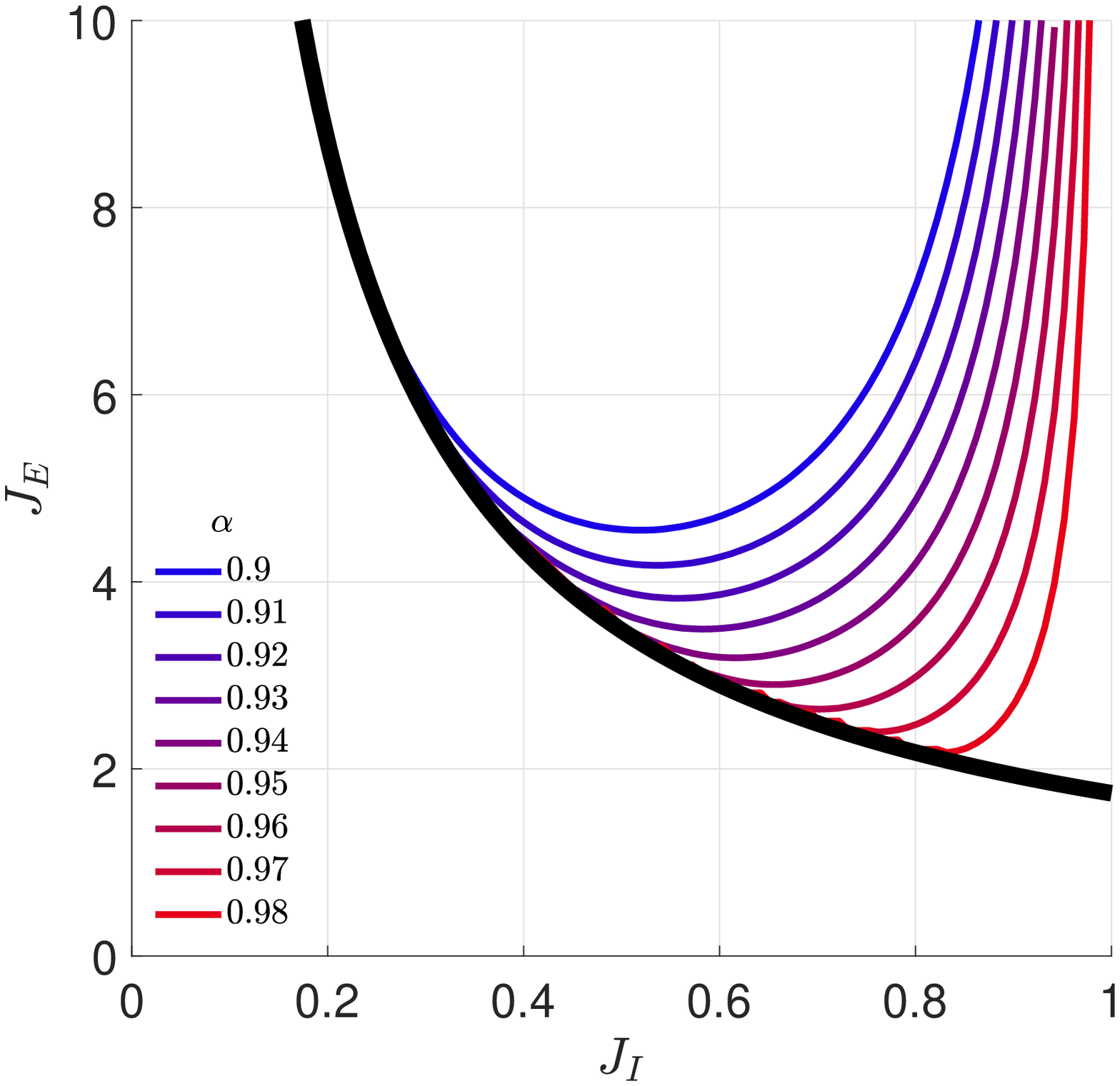}
        \begin{center}
            \caption{\textbf{(c)}  \label{fig:asymmetric_alpha_change}}
        \end{center}
        
    \end{subfigure}\quad \begin{subfigure}[b]{0.3\textwidth}\captionsetup{justification=centering}
        \includegraphics[width=5cm]{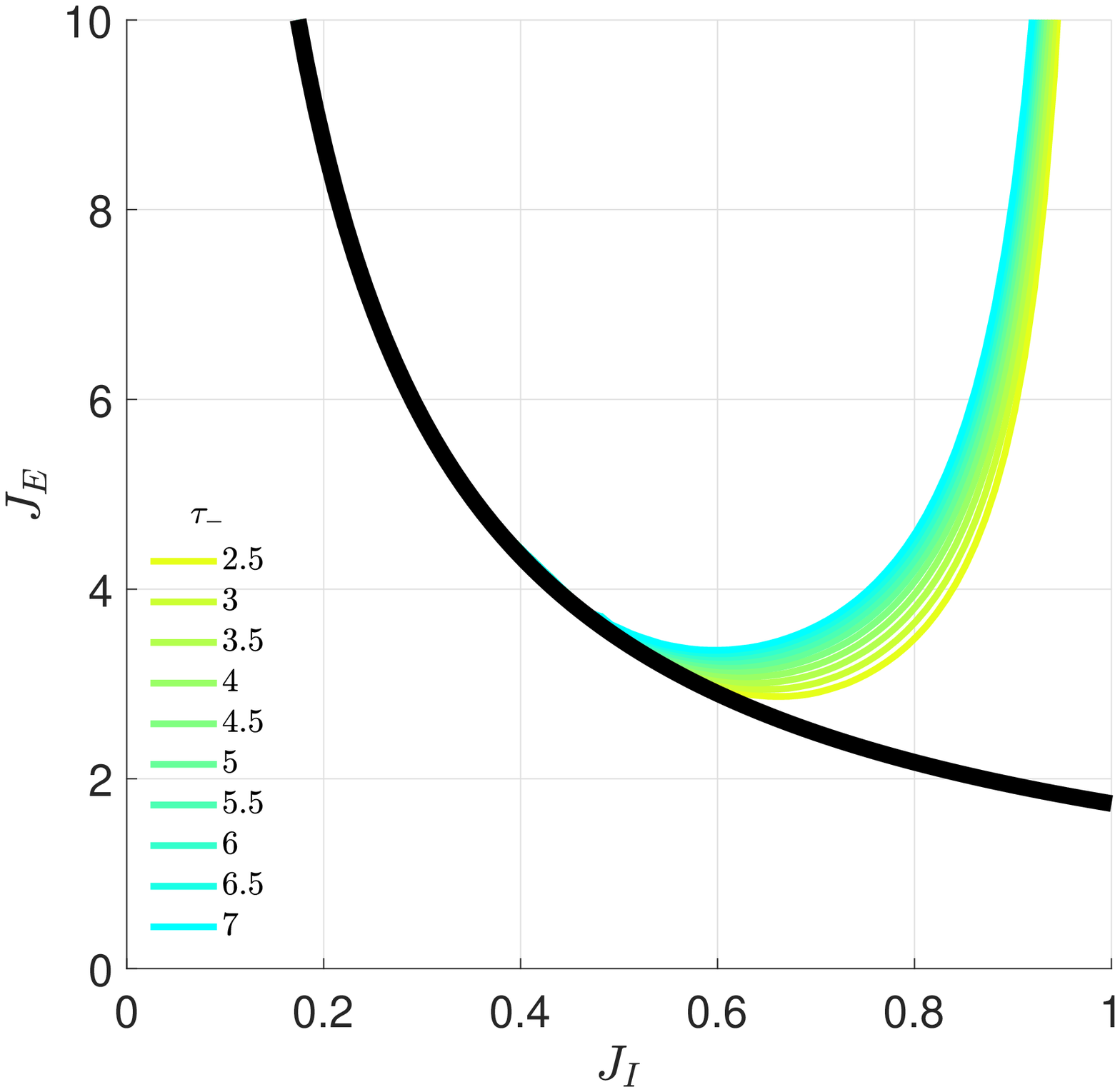}
        \begin{center}
            \caption{\textbf{(d)}  \label{fig:asymmetric_tau_change}}
        \end{center}
        
        \end{subfigure}
    \caption{Nullclines of the STDP dynamics.
    ({\bf \protect\subref{fig:gauss_alpha_change}}) and ({\bf \protect\subref{fig:gauss_tau_change}})
    The nullclines of $J_E$ and $J_I$ for the  difference of Gaussians learning rule (in this case the nullclines are identical for the same choice of parameters $\alpha$, $\tau_\pm$)  
    The nullclines are shown for different values of of $\alpha=0.99, 0.991,...0.999$, differentiated by color, with $\tau_+=2$ and $\tau_-=1$, in (\protect\subref{fig:gauss_alpha_change}).
    The nullclines are shown for different values of $\tau_-=0.25, 0.5,...3.25$, by colors, with $\tau_+=5$ and $\alpha=0.999$, in (\protect\subref{fig:gauss_tau_change}).
    ({\bf \protect\subref{fig:asymmetric_alpha_change}}) and ({\bf \protect\subref{fig:asymmetric_tau_change}}) The nullclines of $J_I$ ($J_E$) for the temporally asymmetric Hebbian (anti-Hebbian) STDP rule.
     The nullclines are shown for different values of $\alpha=0.9, 0.91,...0.99$, with $\tau_+=2$ and $\tau_-=5$, in (\protect\subref{fig:asymmetric_alpha_change}). The nullclines are shown for different values of $\tau_-=2.5, 3,...7$, with $\tau_+=2$ and $\alpha=0.94$, in (\protect\subref{fig:asymmetric_tau_change}). The nullclines were computed using the cosine approximation for the neuronal cross-correlations.}\label{nullclines}.   
\end{figure*}

The difference of Gaussians learning rule, Eq.~(\ref{eq:DOG}), yields
\begin{equation}\label{K_tilde_DOG}
    \tilde{K}_\sigma=\cos\tilde{\varphi}\left(e^{-\frac{\left(\omega\tau_{\sigma,+}\right)^2}{2}}-\alpha e^{-\frac{\left(\omega\tau_{\sigma,-}\right)^2}{2}}\right)
    ,
\end{equation}
with $ \sigma \in \{ E, I \}$. Consequently, the dynamical equations of $J_E$ and $J_I$ will be identical if the characteristic time scales of potentiation and depression are the same; namely, if $\tau_{E,+} = \tau_{I,+}$ and $\tau_{E,-} = \tau_{I,-}$. Note that on the r.h.s.\ of Eq.\ (\ref{K_tilde_DOG}), the term $\cos\tilde{\varphi}$ ensures that $\tilde{K}_\sigma$ is zero on the bifurcation (see Fig.\ \ref{fig:PhiTilde}). Figures \ref{fig:gauss_alpha_change} and \ref{fig:gauss_tau_change} depict the nullclines of $J_E$ and $J_I$ , respectively, for different values of the relative strength of depression, $\alpha$ (in \ref{fig:gauss_alpha_change}), and the characterstic time of depression, $\tau_-$ (in \ref{fig:gauss_tau_change}), differentiated by color. We show that for $\alpha<1$ ($\alpha>1$) and $\tau_+>\tau_-$ ($\tau_+<\tau_-$) the nullcline of $J_E$ and the left branch of the nullcline of $J_I$ are stable (unstable). A fixed point of the STDP dynamics is obtained by the intersection of $J_E$ and $J_I$ nullclines. For the difference of Gaussians rule, a stable fixed point that exhibits rhythmic activity in the gamma band can thus be obtained. However, this requires a delicate adjustment of the parameters characterizing the STDP learning rules.

For the temporally asymmetric Hebbian exponential rule, Eq.~(\ref{eq:TA}) with $H=1$, we obtain
\begin{equation}
\begin{split}
  \tilde{K}_\sigma=&\cos\left(\theta_{\sigma,+}\right)\cos\left(\theta_{\sigma,+}-\tilde{\varphi}_\sigma\right)\\
  &-\alpha \cos\left(\theta_{\sigma,-}\right)\cos\left(\theta_{\sigma,-}+\tilde{\varphi}_\sigma\right)    %
\end{split}
\end{equation}
where $\cos\left(\theta_{\pm}\right)=(1+\left(\omega\tau_\pm\right)^2)^{-1/2}$, $\tilde{\varphi}_E=\tilde{\varphi}$ and $\tilde{\varphi}_I=-\tilde{\varphi}$. Now, due to $\theta_{\sigma,\pm}$,  $\tilde{K}_\sigma$ transitions discontinuously across the bifurcation line, thus inducing discontinuity in $\dot{J}_E$ and $\dot{J}_I$ along the transition from fixed point to the rhythmic region. 
Figures \ref{fig:asymmetric_alpha_change} and \ref{fig:asymmetric_tau_change} depict the nullclines of $J_I$, for the temporally asymmetric Hebbian learning rule for different values of the relative strength of depression, $\alpha$ (in \ref{fig:asymmetric_alpha_change}), and the characteristic time of depression, $\tau_-$ (in \ref{fig:asymmetric_tau_change}), differentiated by color. The left branch of the nullclines of $J_I$ is stable (unstable) for $\alpha<1$ ($\alpha>1$) and $\tau_+<\tau_-$ ($\tau_+>\tau_-$). Interestingly, a considerable part of the $J_I$ nullcline is on the bifurcation line.

For the temporally asymmetric Hebbian learning rule, the dynamics of $J_E$ do not have a nullcline. As a result, a fixed point does not exist, and the temporally asymmetric Hebbian STDP rule cannot stabilize rhythmic activity in the gamma range. 

However, as $\Gamma_{EI}\left(\Delta\right)=\Gamma_{IE}\left(-\Delta\right)$ the temporally asymmetric \emph{anti}-Hebbian exponential rule (Eq.\ (\ref{eq:TA}) with $H=-1$) for excitatory synapses,  $J_E$, in our model, defines the exact same dynamics as that of an inhibitory synapses, $J_I$, with a Hebbian rule (Eq.\ (\ref{eq:TA}) with $H=+1$). Therefore, an asymmetric Hebbian learning rule for $J_I$ and an asymmetric anti-Hebbian learning rule for $J_E$ yield the same nullclines (see Fig.\ \ref{fig:asymmetric_alpha_change}-\ref{fig:asymmetric_tau_change}). Moreover, because a considerable part of the nullcline is on the bifurcation line, no fine tuning of the parameters is required to obtain a fixed point of the STDP dynamics that will generate rhythmic activity at $\omega_d$. Thus, the STDP dynamics have a line attractor on (part of) the bifurcation line.



\section{\label{sec:Discussion}Discussion}
Previous studies have investigated the effects of rhythmic activity on STDP \cite{luz2016oscillations,sherf2020multiplexing,cateau2008interplay,gerstner1996neuronal,gilson2012frequency,karbowski2002synchrony,kerr2013delay,lee2009cortical,masquelier2009oscillations,muller2011spike,pfister2010stdp,luz2016oscillations}. However, in these studies, rhythmic activity was hard-wired in the system and the issue of rhythmogenesis was not addressed.

Rhythmogenesis can be thought of as self organizing temporal activity; i.e., the ability of a non-rhythmic system to spontaneously develop rhythmic activity. In our approach, the process of rhythmogenesis was mapped to a flow on the phase diagram. This mapping relies on the separation of timescales.

Previously, Soloduchin \& Shamir investigated rhythmogenesis using the framework of two neuronal populations with reciprocal inhibition and short term adaptation in the form of firing rate adaptation {\cite{soloduchin2018rhythmogenesis,shamir2019theories}}. The network motif of reciprocal inhibition has been widely reported in the central nervous system {\cite{aksay2007functional,kim2016antagonistic,koyama2018mutual}}. However, it is mainly associated with winner-take-all like competition \cite{jin2002fast,fukai1997simple,shamir2006scaling,hertz1991introduction,bazhenov2013computational,machens2005flexible,zohar2016readout} rather than generating rhythmic activity (but see \cite{danner2017computational,ausborn2018state} in the spinal cord). Here, rhythmogenesis was studied in the framework of a network that is considered a valid hypothesis for generating gamma rhythm in the brain \cite{roxin2006rate,battaglia2007temporal,battaglia2011synchronous}. 

In Soloduchin \& Shamir  {\cite{soloduchin2018rhythmogenesis}}, rhythmogenesis was obtained as a specific stable fixed point on the phase diagram of the system, in which due to the temporal characteristics of the STDP rule, the dynamics of the synaptic weights vanish at a specific frequency. This scenario is similar to the case of temporally symmetric STDP (Fig. \ref{fig:gauss_alpha_change}-\ref{fig:gauss_tau_change}). However, scientifically, this scenario is somewhat disappointing, since we have traded the problem of fine-tuning of the synaptic weights for the problem of fine-tuning of characteristics of the STDP rule {\cite{shamir2019theories}}. 

The temporally asymmetric STDP rule provides a possible solution to the fine-tuning problem of rhythmogenesis, which we term \emph{critical rhythmogenesis} (Fig. \ref{fig:asymmetric_alpha_change}-\ref{fig:asymmetric_tau_change}). Rhythmogenesis in the temporally asymmetric STDP rule is not obtained as a fixed point of the STDP dynamics. Rather, in this case, rhythmogenesis utilizes the discontinuity of $\tilde{\Gamma}$ across the bifurcation line. For a wide range of parameters the flow induced by the STDP is directed from the fixed point region towards the rhythmic region and from the rhythmic region to the fixed point region, and the system will settle on the bifurcation line itself. Consequently, the resultant rhythmic activity will be dictated by bifurcation (e.g., the firing rates will oscillate at $\omega_d$), which is independent of the synaptic plasticity thus accounting for the robustness to the parameters that characterize the STDP. 

Recently, Pernelle and colleagues studied the possible contribution of gap junction plasticity to rhythmic activity {\cite{pernelle2018gap}}. They postulated a plasticity rule for gap junctions that tuned the system to operate on the boundary of asynchronous regular firing activity, on one hand, and rhythmic activity of synchronous bursts, on the other. In the context of our work this can be viewed as an example of critical rhythmogenesis, which explains the robustness of their putative mechanism to variations in the plasticity rule.     

We suggest that the scenario of critical rhythmogenesis may provide a general principle for robustness in biological systems. Assume a certain biological system, which is characterized by set parameters: $\{ x_1, x_2 \ldots x_n\}$ (i.e., $\vec{x}$ is a point in the phase diagram of the system), is required to maintain a certain living condition, $ f(\{ x_1, x_2 \ldots x_n\} ) = 0$. This condition is met by a homeostatic process. The homeostatic process defines the dynamics on $\{ x_1, x_2 \ldots x_n\}$, which are characterized by another set of parameters, $\{ \alpha_1, \alpha_2 \ldots \alpha_m\}$, namely: $ \dot{\vec{x}} = \mathcal{F} (\vec{x}, \vec{\alpha})$. A viable homeostatic process is a choice of parameters, $\vec{\alpha}^*$, such that the homeostatic dynamics will lead the system to a set of parameters, ${x}_{\infty}$, that satisfy the living condition, $ f(\vec{x}_{\infty} ) = 0$. 

How can this be achieved? One possibility is that ${x}_{\infty}$ is a stable fixed point of the homeostatic dynamics, in which $\mathcal{F} (\vec{x}_\infty, \vec{\alpha}^*) = 0$. This solution requires fine-tuning of the parameters that define the homeostatic process, $\vec{\alpha}$. In this case, fluctuations in $\vec{\alpha}$ will generate fluctuations in $\vec{x}$ away from ${x}_{\infty}$. 

An alternative scenario is that the homeostatic
dynamics utilize some discontinuity in the phase diagram. In this scenario the dynamics do not necessarily vanish on ${x}_{\infty}$, $\mathcal{F} (\vec{x}_\infty, \vec{\alpha}^*) \neq 0$. Rather, due to the discontinuity there exists a wide range of parameters, $\{ \alpha_1, \alpha_2 \ldots \alpha_m\}$, such that the dynamics draw the system towards the discontinuity from both sides, as is the case for critical rhythmogenesis. Consequently, this scenario can stabilize the system in a critical condition on the boundary of two phases.  

The idea that the central nervous system may operate in (or near) a critical condition has been suggested in the past, and may have computational advantages \cite{kinouchi2006optimal,shew2011information,shriki2016optimal}. However, this latter scenario also has shortcomings. The most obvious is that it can only be used to ensure and stabilize critical behavior. In addition, it cannot be used when one of the phases near the critical line is lethal. 

An advantage of critical rhythmogenesis is that it allows for rapid switching between rhythmic and non-rhythmic phases, for example by neuromodulators, since the system is on the boundaries of these phases. This raises the question of the likelihood of critical rhythmogenesis: what is the probability that a biological system will `choose' the exact rhythmic activity that also characterizes the bifurcation? One possible explanation is that the opposite took place. In other words, biological systems have evolved to operate at the critical conditions chosen by the critical rhythmogenesis mechanism. Thus, in our example, the characteristic delays, $d$, do not miraculously fit the desired rhythm. Rather, due to the specific values of $d$, the critical rhythmogenesis tunes the system to oscillate at $\omega_d$ which is why the biological system `uses' this specific frequency band. 

In our work we made several simplifying assumption to facilitate the analysis. We studied the dynamics of the effective couplings between excitatory and inhibitory populations and did not incorporate the STDP dynamics of individual synapses. We estimated neuronal correlations using a simplified rate model, and did not study the effects of spiking neurons. These issues are beyond the scope of the current study and will be addressed elsewhere. Nevertheless, this work lays the foundation for studying a novel mechanism for robust homeostatic plasticity, in general, and rhythmogenesis in particular.

\begin{acknowledgments}
This work was supported by the Israel Science Foundation, grant number 300/16.
\end{acknowledgments}

\bibliographystyle{unsrt}
\bibliography{Mybib}

\end{document}